\documentclass[aps,pra,twocolumn]{revtex4-2}%
\usepackage{graphics}
\usepackage{amsmath}
\usepackage{graphicx}
\usepackage{amsfonts}
\usepackage{amssymb}

\begin{document}
\title{Holographic Limitations and Corrections to Quantum Information Protocols}
\author{Stefano Pirandola}
\affiliation{University of York, York YO10 5GH, United Kingdom}


\begin{abstract}
We discuss the limitations imposed on entanglement distribution, quantum teleportation, and quantum communication by holographic bounds, such as the Bekenstein bound and Susskind's spherical entropy bound. For continuous-variable (CV) quantum information, we show how the naive application of holographic corrections disrupts well-established results. These corrections render perfect CV teleportation impossible, preclude uniform convergence in the teleportation simulation of lossy quantum channels, and impose a revised PLOB bound for quantum communication. While these mathematical corrections do not immediately impact practical quantum technologies, they are critical for a deeper theoretical understanding of quantum information theory.
\end{abstract}

\maketitle


\section{Introduction}

Inspired by the holographic principle, the holographic bounds~\cite{BoussoREV} set fundamental limits on the amount of information that can be contained within a given volume of space. Rooted in theories that intersect quantum mechanics, general relativity, and thermodynamics, these bounds suggest that the maximum entropy in a spatial region is directly proportional to its surface area, rather than its volume. Prominent examples include the Bekenstein Bound~\cite{Bekenstein1}, Susskind's spherical entropy bound~\cite{Sus}, and the Bekenstein-Hawking entropy formula for black holes~\cite{BH1,BH2}. These have profound implications for our understanding of gravity, information theory, and the fabric of the universe itself. 

It is very interesting to explore such bounds in the context of quantum information theory. Assuming a simple connection between thermodynamic and von Neumann entropy~\cite{NielsenBOOK}, and assuming that quantum protocols can be operated in the same way no matter if in flat or curved space-time, one can derive simple limitations for the number of qubits that can be entangled or teleported over a certain distance. More interesting, because standard results in continuous-variable (CV) quantum information are related to high-entropy limits, 
the direct application of the holographic bounds to CV protocols leads to fundamental restrictions and corrections. This is the case for CV teleportation~\cite{teleCV,telereview}, its associated tools for quantum channel simulation~\cite{TQCreview,EPJD}, and also for the fundamental limit of quantum communication, known as the Pirandola-Laurenza-Ottaviani-Banchi (PLOB) bound~\cite{PLOB}.   

\section{General scenario}

\begin{figure}[ptbh]
\vspace{-1cm}
\par
\begin{center}
\includegraphics[width=0.48\textwidth] {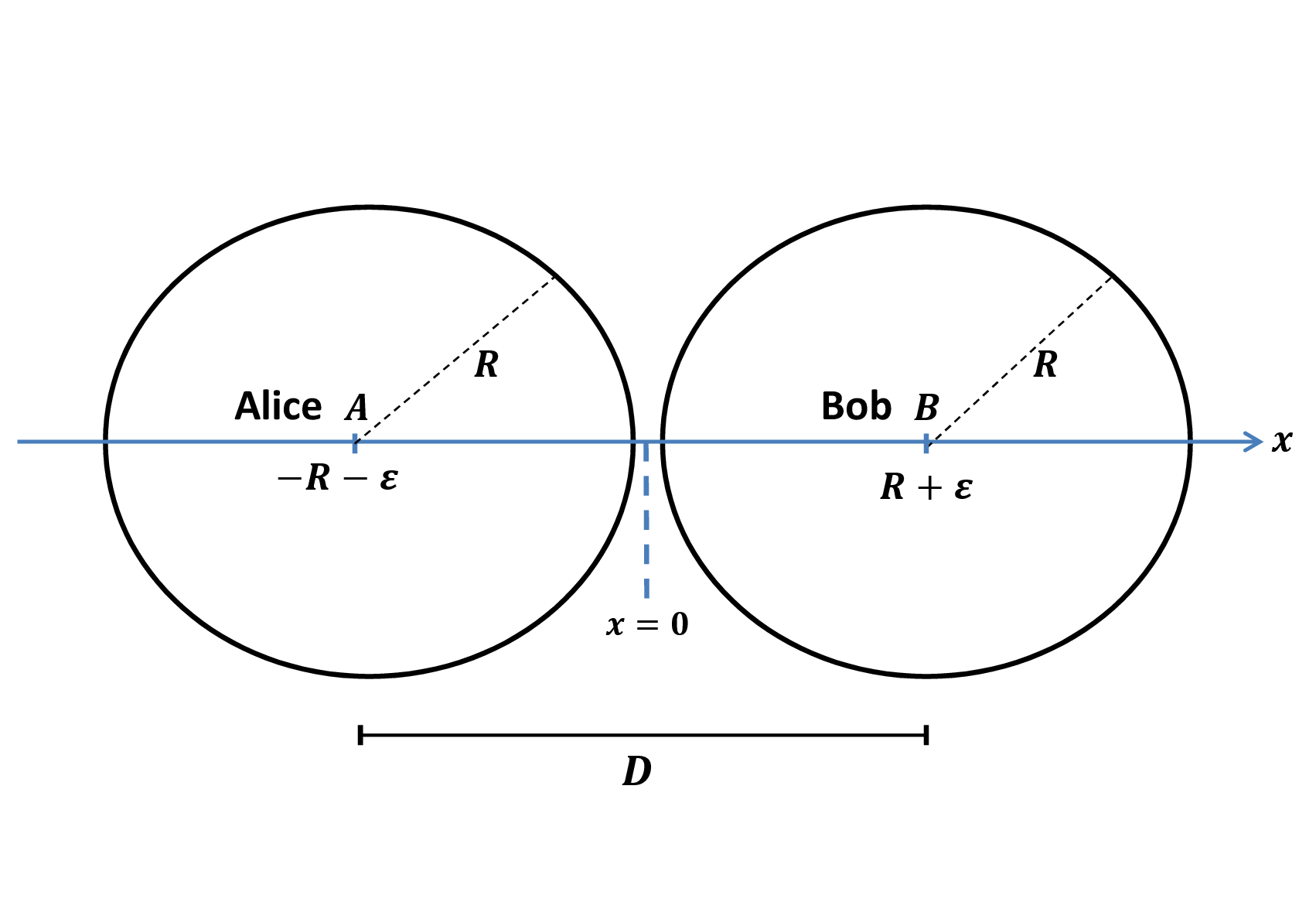}
\end{center}
\par
\vspace{-1.2cm}\caption{Double-sphere scenario. An entanglement
source is located at the origin $x=0$ of the position coordinate
$x$, while Alice is at $-R-\varepsilon$ and Bob at
$R+\varepsilon$, with $\varepsilon$ arbitrarily small. Alice's and
Bob's local labs are within spheres of radius $R$. The middle
source distributes $n$ Bell pairs to the remote parties, separated
by $D=2R+2\varepsilon\simeq2R$. We call
$D$ entanglement or teleportation distance.}
\label{ballsPIC}
\end{figure}

Consider an entanglement source between two parties, Alice and Bob. As shown
in Fig.~\ref{ballsPIC}, the source is located at the origin $x=0$ while Alice
is at position $x=-R-\varepsilon$ and Bob at $x=R+\varepsilon$ with
$\varepsilon \ge 0$ arbitrarily small. Alice-Bob distance is therefore equal to
$D=2R+2\varepsilon \ge 2R$. Suppose that the source distributes
a maximally entangled state $\Psi_{AB}:=|\Psi\rangle_{AB}\langle\Psi|$ where
$|\Psi\rangle_{AB}=d^{-1/2}\sum_{i}|ii\rangle$ with $d$ being the local
dimension of systems $A$ (reaching Alice) and $B$ (reaching Bob).
Because of entanglement distillation, this state is equivalent to
$n=\mathrm{log}_{2}d$ Bell pairs or entanglement bits (ebits), i.e., $n$
copies of the state $|\Psi_{0}\rangle=2^{-1/2}(|00\rangle+|11\rangle)$. This
resource may be used to implement quantum protocols, including teleportation.

Using this shared distilled resource, Alice may teleport an arbitrary
state of $n$ qubits to Bob. She may measure each input qubit (in a
reduced state $\rho $) with the $A$-part of an ebit by performing
a joint Bell detection. The effect of the measurement is to
project Bob's $B$-part of the same ebit onto the state $P_{u}\rho
P_{u}^{\dagger}$, where $\{P_{u}\}_{u=0}^{3}=\{I,X,Y,Z\}$ is the
set of Pauli operators~\cite{NielsenBOOK} plus the identity. Then,
Alice transmits the 2-bit value $u$ to Bob. Thanks to this
classical communication (CC), Bob
can undo the unitary $P_{u}$ from his $B$ qubit, thus
reconstructing the state of Alice's input qubit. This procedure
can be repeated for all the $n$ input qubits, so that Alice's global
state is perfectly transferred to Bob's qubits. For this transfer,
$n$ ebits are consumed and $2n$ classical bits need to be
communicated. It is natural to assume that input qubits and shared
ebits are of the same nature (e.g., same
mass). Here we refer to $D\simeq2R$ as to entanglement or
teleportation distance.

Because the non-local state $\Psi_{AB}$ is pure, it has von
Neumann entropy $S_{AB}=0$. At the same time, the reduced local
states of Alice and Bob are maximally mixed, i.e.,
$\rho_{A}=\rho_{B}=d^{-1}I$ where $I$ is the identity operator.
Thus, in Alice's and Bob's labs, the local systems have maximal
von Neumann entropy $S_{A}=S_{B}=n$ qubits. This is also known as
entanglement entropy.

For the following derivations, we make the following naive assumptions:
\begin{itemize}
\item The von Neumann entropy of a subsystem provides its
thermodynamic entropy. More precisely, we assume that the
thermodynamic entropy is obtained by changing the log-base to nats
and including the Boltzmann constant $k$, so that
$S_{A}^{\mathrm{th}}=S_{B}^{\mathrm{th}}=kn\ln2$.


\item Quantum protocols for entanglement distribution, teleportation, and quantum communication are assumed to be operated in the same way no matter if they are in flat or curved space-time.

\end{itemize}

\section{Holographic bounds}

Given the assumptions above one can easily show the immediate and general limitations that holographic bounds impose on protocols for entanglement distribution and quantum teleportation. These limitations are expressed in terms of the number of ebits that can be distributed or qubits that can be teleported over some distance $D$. We discuss these limitations from the perspective of the Bekenstein bound~\cite{Bekenstein1}, and we then extend the discussion to Susskind's spherical bound~\cite{Sus}.

\subsection{Holographic limits from the Bekenstein bound}

The Bekenstein bound holds for any weakly gravitating matter system in an asymptotically
flat space-time. In such an approximately flat scenario, consider a single sphere of
radius $R$ centered on Alice's local system $A$ (e.g., see the left sphere of Fig.~\ref{ballsPIC}). The thermodynamic entropy of Alice's system must satisfy the bound~\cite{Bekenstein1}
\begin{equation}
S_{A}^{\mathrm{th}}(=kn\ln2)\leq\frac{2\pi kRE}{\hbar c}, \label{bekenstein2}%
\end{equation}
where $E$ is the energy within the sphere, $\hbar$ is the Planck constant, and
$c$ is the speed of light in vacuum. It is easy to see that this inequality provides a macroscopic 
upper bound to the number $n$ of qubits that can be compressed within a sphere of radius $R$
and internal energy $E$.

Because $D \ge 2R$,\ we may write
\begin{equation}
DE\geq\frac{\hbar c\ln2}{\pi}n\simeq 6.97 \times 10^{-27}n~~~\mathrm{[J \cdot m]}. \label{energy2}%
\end{equation}
This can be seen as an upper bound on the number of ebits
that can be shared by two parties that are separated by $D$ and
have local energy $E$ (as depicted in the two-sphere scenario of Fig.~\ref{ballsPIC}). In turn, this is also 
an upper bound on the number of qubits that can be teleported over distance $D$ by consuming an entanglement
source with total energy $2E$~\cite{teleNOTE}.

Now assume that the local systems $A$ and $B$ in Fig.~\ref{ballsPIC} are at rest, so that their local
mass $M$ provides an equivalent energy of $E=Mc^{2}$ (we assume there is no
charge or momentum associated with the qubits, so we also ignore relativistic effects). Then, Eq.~(\ref{energy2})
becomes%
\begin{equation}
DM\geq\frac{\hbar\ln2}{\pi c}n\simeq 7.77 \times 10^{-44}n ~~~\mathrm{[m \cdot Kg/s]}, \label{eqmain}%
\end{equation}
which provides a macroscopic bound to the number $n$ of qubits
that can be entangled at distance $D$ with local mass $M$. If we
assume that $M$ is also the mass of an object to be teleported (in
a mass-preserving teleportation), then Eq.~(\ref{eqmain}) bounds
the number $n$ of massive qubits that can be teleported over distance $D$.
It is clear that, fixing the number $n$ of qubits in
Eq.~(\ref{eqmain}), we get a trade-off between entanglement/teleportation distance $D$ and local mass $M$. Surprisingly, the
lesser $M$ is, the greater $D$ needs to be, so that we can
entangle or teleport massive objects only beyond a minimum
distance, which becomes infinite in the limit of $M\rightarrow0$.

\subsection{Limitations from the spherical entropy bound}

The violation of the Bekenstein bound is associated with a
violation of the second law of thermodynamics. It is generally
considered to be valid for weakly gravitating systems in spherical
symmetry~\cite{BoussoREV}. Under the same kind of symmetry, one may
consider Susskind's spherical entropy bound~\cite{Sus} which can
be extended to strongly gravitating systems, i.e., truly curved
space-time.

The thermodynamic entropy in a volume of space bounded by a spherical surface
with area $\mathcal{A}=4\pi R^{2}$ and radius $R$ must satisfy the bound%
\begin{equation}
S^{\mathrm{th}}\leq\frac{k\mathcal{A}}{4l_{p}^{2}}=\frac{\pi kR^{2}}{l_{p}%
^{2}}, \label{spherical}%
\end{equation}
where $l_{p}=\sqrt{G\hbar/c^{3}}\simeq1.6\times10^{-35}$m is the Planck
length, with $G$ being the gravitational constant. The saturation of this
bound is achieved by the most entropic possible object, a black hole. This
extremal value is also known as Bekenstein-Hawking (BH) entropy
$S_{\mathrm{BH}}$, in which case $R$ and $\mathcal{A}$ are radius and area of
the event horizon, respectively. In the absence of charge and angular
momentum, we may consider a Schwarzschild black hole with mass $M$ whose event
horizon has radius $R_{\text{S}}=2GM/c^{2}$. Then the BH entropy becomes
\begin{equation}
S_{\mathrm{BH}}=\frac{\pi kR_{\text{S}}^{2}}{l_{p}^{2}}=\frac{2\pi
kcR_{\text{S}}M}{\hbar}. \label{BHeq}%
\end{equation}

In our entanglement distribution scenario (see
Fig.~\ref{ballsPIC}), Alice's and Bob's local systems have
$S_{A}^{\mathrm{th}}=S_{B}^{\mathrm{th}}=kn \ln2$, so the entropy
bound in Eq.~(\ref{spherical}) directly provides
\begin{equation}
n\leq
\frac{\pi}{\ln2}(R/l_{p})^{2},~~\frac{D}{l_{p}}\geq\sqrt{\frac{4
\ln2}{\pi}n}~~.
\label{minDfff}%
\end{equation}
Thus, \textit{independently} of their mass, the maximum number $n$
of qubits that can be entangled or teleported between Alice and
Bob is limited by the area of the circle $\pi R^{2}$ in Planck
units. Similarly, there is a universal minimum distance $D/l_{p}$
for $n$-qubit entanglement and teleportation that scales
quadratically in the number of qubits~\cite{NOTA_pre}.

Note that tighter restrictions can be derived from the `t Hooft bound for
ordinary matter~\cite{thooft1,thooft2}. If we exclude energies leading to
gravitational collapse and assume an approximately flat spacetime, the maximum
thermodynamic entropy within a sphere with area $\mathcal{A}$ scales as
$S^{\mathrm{th}}\lesssim k(\mathcal{A}/l_{p}^{2})^{3/4}$. Replacing
$S^{\mathrm{th}}=kn \ln2$, we then find%
\begin{equation}
n\lesssim(R/l_{p})^{3/2},~~\frac{D}{l_{p}}\gtrsim n^{2/3}.
\end{equation}
These scalings are clearly tighter than those in Eq. (\ref{minDfff}).

\subsection{Black-hole creation dynamics}

For the sake of completeness, let us describe the previous process by setting equality in
Eq.~(\ref{eqmain}), so that $D_{\text{min}}M=n\hbar \ln2/(\pi c)$.
This is equivalent to say that we are considering the smallest
spheres capable of enclosing Alice's and Bob's systems. By
increasing the local mass $M$, the entanglement/teleportation distance
$D_{\text{min}}$ decreases. Simultaneously, the Schwarzschild
radius
associated with Alice's and Bob's systems increases as $R_{\text{S}}%
=2GM/c^{2}$. At the critical point $D_{\text{min}}=2R_{\text{S}}$,
Alice's and Bob's systems become a pair of entangled black holes
whose event horizons are tangent. Clearly, teleportation can no
longer work because the CC needed to perform the protocol cannot
escape the horizons. 

One can easily check that, at the critical
point, the local
masses must be equal to%
\begin{equation}
\frac{{M}}{{m_{p}}}=\sqrt{\frac{n \ln2}{4\pi}},
\end{equation}
where $m_{p}=\sqrt{\hbar c/G}\simeq21.76$$\mu$g is the Planck mass.
It is then easy to see that the minimum distance is
$D_{\text{min}}/l_{p}=\sqrt{4n \ln2/\pi}$, which saturates the
bound in Eq.~(\ref{minDfff}). For $D>D_{\text{min}}$, $n$-qubit
protocols for entanglement distribution and teleportation 
become in principle possible because Alice's and Bob's
labs can be located outside the event horizons within their local
spheres.

\section{Holographic limitations to CV protocols}

Let us now show what type of implications and limitations the spherical bound 
would have for continuous-variable (CV) quantum information. This is a field where a number of fundamental
results are obtained by taking limits for infinite entanglement,
squeezing or modulation, which means that these limits imply
infinite entropy. 

\subsection{Limits to Gaussian modulation and two-mode squeezing}


Let us start from the problem of signal modulation. In a typical CV\ protocol for quantum communication,
Alice prepares an input alphabet of coherent states whose
amplitude $\alpha$ is
modulated according to a Gaussian distribution. On average this is a 
thermal state with $\bar{n}$ mean number of
photons and entropy
\begin{equation}
h(\bar{n}):=(\bar{n}+1)\log_{2}(\bar{n}+1)-\bar{n}\log_{2}\bar{n}.
\end{equation}

It is easy to show that $h(\bar{n})\geq\log_{2}(e\bar{n})$ so that,
multiplying by $\ln2$ and including the Boltzmann constant $k$, we compute
Alice's thermodynamic entropy\ $S_{A}^{\mathrm{th}}\geq k\ln(e\bar{n})$. Using
the latter inequality in the spherical bound of Eq.~(\ref{spherical}) leads to%
\begin{equation}
\bar{n}\leq e^{\pi R_{p}^{2}-1}, \label{limitCV}%
\end{equation}
where $R_{p}:=R/l_{p}$ is the radius in Planck units. The first immediate
consequence of Eq.~(\ref{limitCV}) is that Alice's signal alphabet must be
limited by her radius, so that infinite modulation is not allowed for any
finite $R_{p}$.

The same bound clearly holds for the maximal amount of CV
entanglement that can be shared by Alice and Bob. The fundamental
state to consider here is the two-mode squeezed vacuum (TMSV)
state $\Phi_{AB}^{\mu}$ with parameter
$\mu=\bar{n}+1/2$. Tracing out Bob's $B$-part of this
state provides Alice with a local thermal state
$\rho_{A}^{\bar{n}}$ with $\bar{n}=\mu-1/2$ mean photons.
Therefore Alice's reduced state has thermodynamic entropy
$S_{A}^{\mathrm{th}}\geq k\ln(e\bar{n})$ and the spherical bound
provides Eq.~(\ref{limitCV}) or, equivalently,
\begin{equation}
\mu\leq\mu_{\max}(R_{p}):=e^{\pi R_{p}^{2}-1}+1/2.\label{maxmu}%
\end{equation}
As we will see below, the bound in Eq.~(\ref{maxmu}) has drastic
consequences for CV\ teleportation~\cite{teleCV,telereview} and the teleportation
simulation of channels~\cite{PLOB}. It will also impose holographic corrections
to the quantum capacities of bosonic channels~\cite{NOTA}.

\subsection{Holographic no-go for ideal CV\ teleportation}

Let us apply the CV teleportation protocol to mode $a$ of an input TMSV
state $\Phi_{Ca}^{\tilde{\mu}}$ by using another TMSV state
$\Phi_{AB}^{\mu}$ as a resource. The ideal CV\ Bell detection on
modes $a$ and $A$, and the CC of the outcome realizes an
approximate identity channel $\mathcal{I}_{a\rightarrow B}^{\mu}$
from mode $a$ to mode $B$. This is strongly (i.e., point-wise)
equivalent to an additive-noise Gaussian channel with added
noise~\cite{oldREV,GerLimited,RicFINITE,NOTAcomp}
\begin{equation}
\xi=2\mu-\sqrt{4\mu^{2}-1}. \label{added}%
\end{equation}
When applied to $\Phi_{Ca}^{\tilde{\mu}}$, we get the output $\Phi_{CB}^{\mu,\tilde{\mu}}:=\mathcal{I}_{C}\otimes\mathcal{I}_{a\rightarrow B}^{\mu
}(\Phi_{Ca}^{\tilde{\mu}})$ where $\mathcal{I}_{C}$ is the identity channel applied to system $C$. Consider the (square-root) quantum fidelity
\begin{equation}
F(\mu,\tilde{\mu})=\left\Vert \sqrt{\Phi_{Ca}^{\tilde{\mu}}}\sqrt{\Phi
_{CB}^{\mu,\tilde{\mu}}}\right\Vert _{1}%
\end{equation}
between the input and the output (teleported) state, where $||O||_{1}%
:=\mathrm{Tr}\sqrt{O^{\dagger}O}$ is the trace norm. In the present case, this is the fidelity of teleporting CV entanglement~\cite{Schum}. Using the formula for Gaussian
states~\cite{banchiPRL2015}, we explicitly compute (see also Ref.~\cite[App.~A]{TQCreview})
\begin{equation}
F(\mu,\tilde{\mu})=\left\{  1-4\tilde{\mu}\left[  \sqrt{4\mu^{2}-1}+\tilde
{\mu}-2\mu\left(  1+2\tilde{\mu}\xi\right)  \right]  \right\}  ^{-1/4}.
\label{fid}%
\end{equation}

This expression would go to 1 if we could take the limit of $\mu
\rightarrow+\infty$ but, unfortunately, we have the holographic bound $\mu\leq\mu_{\max}(R_{p})$
of Eq.~\eqref{maxmu} for any radius $R_{p}$. This implies $F<1$ for any finite $R_{p}$,
so we cannot perfectly teleport CV entanglement.
The maximum fidelity in Eq.~(\ref{fid}) is achieved for
$\tilde{\mu}=1/2$, corresponding to the teleportation of the
vacuum state from $a$ to $B$. At Planckian distance $R_{p}=1$, we have
maximal fidelity $F\simeq0.986403$. At larger $R_{p}$, the
fidelity approaches $1$ but remains $<1$ at any finite radius.

This reasoning implies that CV\ teleportation would \textit{not} strongly
converge to the identity channel unless Alice and Bob are
separated by an infinite distance. Note that this problem affects
not only the strong convergence but also the uniform convergence of CV teleportation~\cite{TQCreview,EPJD}. 

\subsection{No uniform convergence in teleportation simulation}

Because the holographic bounds implies that CV\ teleportation
cannot be perfect at finite distance, we have that many
applications of this tool are also affected. This includes the
teleportation simulation of bosonic Gaussian channels. To
illustrate the idea consider the bosonic pure-loss channel which
is the most relevant Gaussian channel. This may be
represented by a beam splitter with transmissivity $0<\eta<1$
which mixes an incoming bosonic mode with an environmental vacuum
mode.

Because the pure-loss channel $\mathcal{E}_{\eta}$ is teleportation
covariant~\cite{PLOB}, it may be simulated by using the CV\ teleportation
protocol $\mathcal{T}$ implemented over its quasi-Choi matrix $\chi_{\eta
}^{\mu}:=\mathcal{I}_{A}\otimes\mathcal{E}_{\eta}(\Phi_{AB}^{\mu})$. In other
words, we may write the simulation channel $\mathcal{E}_{\eta}^{\mu}%
(\rho):=\mathcal{T}(\rho\otimes\chi_{\eta}^{\mu})$ for any input state $\rho$.
One can check that $\mathcal{E}_{\eta}^{\mu}=\mathcal{E}_{\eta}\circ
\mathcal{I}^{\mu}$ and write
\begin{equation}
\lim_{\mu\rightarrow\infty}\left\Vert \mathcal{E}_{\eta}-\mathcal{E}_{\eta
}^{\mu}\right\Vert _{\diamond}=0. \label{claim}%
\end{equation}
Recall that the diamond distance is the appropriate distance
between channels and defined by the following optimization of the trace
distance over bipartite states $\rho_{AB}$
\begin{equation}
\left\Vert \mathcal{E}_{\eta}-\mathcal{E}_{\eta}^{\mu}\right\Vert _{\diamond
}:=\sup_{\rho_{AB}}\left\Vert \mathcal{I}_{A}\otimes\mathcal{E}_{\eta}(\rho_{AB}
)-\mathcal{I}_{A}\otimes\mathcal{E}_{\eta}^{\mu}(\rho_{AB})\right\Vert _{1}.
\label{defDIAM}%
\end{equation}

Holographically, we cannot take the limit in Eq.~(\ref{claim}) due to
Eq.~(\ref{maxmu}), so that the uniform convergence is excluded by a non-zero
lower bound. In fact, we have
\begin{align}
&  \left\Vert \mathcal{E}_{\eta}-\mathcal{E}_{\eta}^{\mu}\right\Vert
_{\diamond}\overset{(1)}{\geq}\left\Vert \left\vert 00\right\rangle
\left\langle 00\right\vert -\left\vert 0\right\rangle \left\langle
0\right\vert \otimes\mathcal{E}_{\eta}^{\mu}(\left\vert 0\right\rangle
\left\langle 0\right\vert )\right\Vert _{1}\nonumber\\
&  \overset{(2)}{\geq}2[1-\left\langle 0\right\vert \mathcal{E}_{\eta}^{\mu
}(\left\vert 0\right\rangle \left\langle 0\right\vert )\left\vert
0\right\rangle ]\nonumber\\
&  \overset{(3)}{=}2\eta\xi(1+\eta\xi)^{-1}, \label{LBdia}%
\end{align}
where: (1)~we pick a particular state (the vacuum $\rho_{AB}=\rho_{00}:=\left\vert
00\right\rangle \left\langle 00\right\vert $) in the optimization in
Eq.~(\ref{defDIAM}) and use the fact that $\mathcal{I}(\rho_{00}%
)=\mathcal{E}_{\eta}(\rho_{00})=\rho_{00}$; (2)~we use the quantum Chernoff
bound~\cite{QCB} $\left\Vert \rho-\sigma\right\Vert _{1}\geq2[1-C(\rho
,\sigma)]$ where $C(\rho,\sigma):=\inf_{s\in\lbrack0,1]}\mathrm{Tr}(\rho
^{s}\sigma^{1-s})$ for any pair of states $\rho$ and $\sigma$, and the fact
that, for pure $\rho:=\left\vert \varphi\right\rangle \left\langle
\varphi\right\vert $, we may write $C(\left\vert \varphi\right\rangle
,\sigma)=F(\left\vert \varphi\right\rangle ,\sigma)^{2}=\left\langle
\varphi\right\vert \sigma\left\vert \varphi\right\rangle $. Finally, in (3) we
use the fact that $\mathcal{E}_{\eta}^{\mu}(\left\vert 0\right\rangle
\left\langle 0\right\vert )$ is a thermal state with variance $\eta\xi+1/2$ and we apply the formula of the fidelity between Gaussian states~\cite{banchiPRL2015}.

The tighter lower bound for the diamond distance $\left\Vert
\mathcal{E}_{\eta}-\mathcal{E}_{\eta}^{\mu}\right\Vert
_{\diamond}$ is obtained when Eq.~(\ref{maxmu}) is saturated,
i.e., for $\mu=\mu_{\max}(R_{p})$. Correspondingly, the added
noise $\xi$ in Eq.~\eqref{added} takes the minimum possible value
\begin{equation}
\xi_{\min}(R_{p})=2\mu_{\max}-\sqrt{4\mu_{\max}^{2}-1}. \label{addedMINvalue}
\end{equation}
However, for any finite $R_{p}$, we have $\xi_{\min}(R_{p})>0$ so that
\begin{gather}
\min_{\mu}\left\Vert \mathcal{E}_{\eta}-\mathcal{E}_{\eta}^{\mu}\right\Vert
_{\diamond}=\left\Vert \mathcal{E}_{\eta}-\mathcal{E}_{\eta}^{\mu_{\max}%
}\right\Vert _{\diamond}\nonumber\\
\geq2\eta\xi_{\min}(R_{p})[1+\eta\xi_{\min}(R_{p})]^{-1}>0~.
\end{gather}
For instance, for $\eta=1/2$ and Planckian radius $R_{p}=1$, we have
$\left\Vert \mathcal{E}_{\eta}-\mathcal{E}_{\eta}^{\mu_{\max}}\right\Vert
_{\diamond}>0.0273791$. For large $R_{p}$, we expand the lower bound at the
leading order and write%
\begin{equation}
\left\Vert \mathcal{E}_{\eta}-\mathcal{E}_{\eta}^{\mu_{\max}}\right\Vert
_{\diamond}\geq\frac{\eta}{2}e^{-\pi R_{p}^{2}}>0,
\end{equation}
so the uniform convergence to $\mathcal{E}_{\eta}$ is not possible.

Similarly, we may enforce an holographic upper bound to the diamond distance.
In fact, we may compute~\cite{NOTAcomp2}
\begin{align}
\left\Vert \mathcal{E}_{\eta}-\mathcal{E}_{\eta}^{\mu}\right\Vert _{\diamond}
&  \leq\delta:=2\sqrt{\frac{\eta\xi}{\eta\xi+1-\eta}}\label{deltaD}\\
&  \simeq\sqrt{\frac{\eta}{(1-\eta)\mu}}+O(\mu^{-3/2}).
\end{align}
Using Eq.~(\ref{maxmu}), for large $R_{p}$ we then derive
\begin{equation}
\left\Vert \mathcal{E}_{\eta}-\mathcal{E}_{\eta}^{\mu_{\max}}\right\Vert
_{\diamond}\lesssim\sqrt{\frac{\eta}{1-\eta}}e^{-\frac{\pi}{2}R_{p}^{2}}.
\end{equation}
As we see below, this bound imposes conditions on the quantum
capacities of the pure-loss channel.

\section{Holographic corrections to the quantum communication limit}




The ultimate performance for quantum key distribution (QKD), entanglement distribution, and quantum state transmission over a pure-loss channel is provided by the PLOB bound~\cite{PLOB}. More precisely, since the PLOB (upper) bound coincides with the lower bound proven in Ref.~\cite{LB2009}, it automatically establishes several capacities for the lossy channel $\mathcal{E}_{\eta}$. It shows that $K(\mathcal{E}_{\eta})=D_2(\mathcal{E}_{\eta})=Q_{2}(\mathcal{E}_{\eta})=-\log_{2}(1-\eta)$, where $K$ is the secret key capacity, $D_{2}$ is the two-way assisted entanglement distribution capacity, and $Q_{2}$ is the two-assisted quantum capacity.
All these capacities are generally assumed to be assisted by two-way classical communication. 

One of the techniques used to prove the PLOB bound is teleportation
stretching, where the tool of quantum channel simulation is used to re-organize the most general
possible two-assisted adaptive quantum protocol into a simpler block version, with no need for feedback
CC between the remote parties (see Ref.~\cite{TQCreview}\ for a general
review). In the bosonic setting, the optimal simulation of a channel via
teleportation requires taking the limit for infinite CV entanglement in the
resource state. Because this limit cannot be taken according to Eq.~(\ref{maxmu}), we need to compute holographic corrections.

Consider Alice and Bob, each within a radius $R_{p}$ and connected by
pure-loss channel $\mathcal{E}_{\eta}$. Assume that they implement the most
general $(N,\mathcal{R},\varepsilon)$-protocol for entanglement distribution.
This means that they use the channel $N$ times interleaved with local
operations (LOs) and two-way CCs, and finally share a bipartite state
$\rho_{N}$ which is $\varepsilon$-close (in trace distance) to a tensor
product $\phi^{\otimes N\mathcal{R}}$ of $N\mathcal{R}$ Bell pairs. Due to the spherical
bound, we must have Eq.~(\ref{minDfff}) and therefore
\begin{equation}
\mathcal{R}\leq \pi R_{p}^{2}/(N \ln2).\label{Rlarge}%
\end{equation}
At macroscopic distances, $R_{p}$ is\ so extremely large that the bound in Eq.~(\ref{Rlarge}) is very large even with $N\gg1$.

To compute a tighter upper bound, we replace each instance of the channel $\mathcal{E}_{\eta}$ with its
simulation $\mathcal{E}_{\eta}^{\mu}(\rho)=\mathcal{T}(\rho\otimes\chi_{\eta}^{\mu
})$ for some LOCC $\mathcal{T}$, where $\chi_{\eta}^{\mu}$ is the channel's quasi-Choi matrix. 
This operation generates a simulated protocol with output $\rho_{N}^{\mu
}$ such that $\left\Vert \rho_{N}^{\mu}-\rho_{N}\right\Vert
_{1}\leq N\delta$ where $\delta$ is the bound defined in Eq.~(\ref{deltaD})~\cite{NOTApeeling}.
Using the triangle inequality, we get
\begin{align}
\left\Vert \rho_{N}^{\mu}-\phi^{\otimes N\mathcal{R}}\right\Vert _{1} &
\leq\left\Vert \rho_{N}^{\mu}-\rho_{N}\right\Vert _{1}+\left\Vert \rho
_{N}-\phi^{\otimes N\mathcal{R}}\right\Vert _{1}\nonumber\\
&  \le \varepsilon+N\delta:=\tilde{\varepsilon}~.
\end{align}
Assuming the condition $\tilde{\varepsilon}\leq1/2$\ we may write a Fannes-type inequality for the relative entropy of
entanglement (REE)~\cite{REE1,REE2} $E_{\text{R}}$. Following Ref.~\cite{PLOB}, this is given by 
\begin{equation}
E_{\text{R}}(\phi^{\otimes N\mathcal{R}})\leq E_{\text{R}}(\rho_{N}^{\mu
})+4\tilde{\varepsilon}\log_{2}d+2H_{2}(\tilde{\varepsilon}),\label{tobecome}%
\end{equation}
where $d=2^{2N\mathcal{R}}$ is the total dimension of the target state
$\phi^{\otimes N\mathcal{R}}$ and $H_{2}$ is the binary Shannon
entropy.

Because the REE bounds the two-way distillable entanglement of a quantum
state, we may write $N\mathcal{R}\leq E_{\text{R}}(\phi^{\otimes N\mathcal{R}%
})$. Then, because the channel $\mathcal{E}_{\eta}^{\mu}$ is simulated by the
resource state $\chi_{\eta}^{\mu}$, we may apply the stretching technique from
Ref.~\cite{PLOB} and decompose the output state as $\rho_{N}^{\mu}%
=\Lambda(\chi_{\eta}^{\mu\otimes N})$ for a trace-preserving LOCC $\Lambda$.
Finally, because the REE\ is monotonic under $\Lambda$ and multiplicative over
tensor products, we may write $E_{\text{R}}(\rho_{N}^{\mu})\leq NE_{\text{R}%
}(\chi_{\eta}^{\mu})$. Therefore, by employing all these considerations, we
have that Eq.~(\ref{tobecome}) becomes
\begin{equation}
\mathcal{R}\leq\frac{E_{\text{R}}(\chi_{\eta}^{\mu})+2N^{-1}H_{2}%
(\tilde{\varepsilon})}{1-8\tilde{\varepsilon}}.\label{Rgen}%
\end{equation}

For a given radius $R_{p}$, we can take the maximum value $\mu=\mu_{\max
}(R_{p})$, so that we get 
\begin{equation}
\tilde{\varepsilon}=\tilde{\varepsilon}_{\min}:=\varepsilon+2N\sqrt{\frac
{\eta\xi_{\min}(R_{p})}{\eta\xi_{\min}(R_{p})+1-\eta}},
\end{equation}
where we have used Eqs.~\eqref{addedMINvalue} and~\eqref{deltaD}. By replacing $\mu_{\max}$ and $\tilde{\varepsilon}_{\min}$ in Eq.~(\ref{Rgen}%
), we get a bound for the optimal rate of an $(N,\mathcal{R},\varepsilon
)$-protocol implemented at distance $2R_{p}$ over a pure-loss channel with
transmissivity $\eta$. 

Take the limits for large $\mu\simeq e^{\pi R_{p}^{2}}$ and small $\varepsilon$, so
that
\begin{equation}
\tilde{\varepsilon}\simeq\varepsilon+N\sqrt{\frac{\eta}{1-\eta}}%
e^{-\frac{\pi}{2}R_{p}^{2}}\simeq0.\label{epsEXP}%
\end{equation}
Let us expand $(1-8\tilde{\varepsilon})^{-1}\simeq1+8\tilde{\varepsilon
}+O(\tilde{\varepsilon}^{2})$ and $H_{2}(\tilde{\varepsilon})\simeq
\tilde{\varepsilon}/\ln2+O(\tilde{\varepsilon}\ln\tilde{\varepsilon})$, so
that Eq.~(\ref{Rgen}) becomes
\begin{equation}
\mathcal{R}\leq(1+8\tilde{\varepsilon})E_{\text{R}}(\chi_{\eta}^{\mu
})+O\left(  \tilde{\varepsilon}^{2},\tilde{\varepsilon}N^{-1}\right)
.\label{tobecome2}%
\end{equation}
At the leading order in $\mu$ we may also write~\cite{PLOB}
\begin{equation}
E_{\text{R}}(\chi^{\mu}_{\eta}) \lesssim -\log_{2}(1-\eta)+O(\mu^{-1}).\label{erEXP}%
\end{equation}
By using Eqs.~(\ref{epsEXP}) and~(\ref{erEXP}) in Eq.~(\ref{tobecome2}), we
derive the following modified version of the PLOB bound%
\begin{equation}
\mathcal{R}\lesssim\left(  1+8\varepsilon+8N\sqrt{\frac{\eta}{1-\eta}%
}e^{-\frac{\pi}{2}R_{p}^{2}}\right)  [-\log_{2}(1-\eta)],\label{erre}%
\end{equation}
where we see the holographic correction due to $R_{p}$. 

This minor adjustment may not hold immediate practical significance for quantum technologies. However, it suggests that the PLOB upper bound might not be precisely aligned with a corresponding lower bound. Further work is needed 
in this direction, specifically in terms of extending the coherent~\cite{Coh1,Coh2} and reverse coherent~\cite{LB2009} information to include holographic corrections.

\section{Conclusions}
We have investigated the implications of directly imposing holographic bounds on the processes of entanglement 
distribution and quantum teleportation. In the first general discussion, we over-viewed how
these bounds would provide direct constraints to the maximum number of ebits and qubits 
that can be involved in these quantum protocols, together with limitations on the minimum distance
for entanglement distribution or teleportation. More interestingly, we have analyzed the 
effects of holography on continuous-variable quantum information, where results are
typically achieved in the limit of unbounded entropy. In this case, we 
have explored how standard results would break down, such as ideal CV teleportation, while others would need corrections, such as the PLOB bound for quantum communication.

\section*{Acknowledgements}
This work was supported by the EPSRC via the UK Quantum Communications Hub with Grants No.
EP/M013472/1 and No. EP/T001011/1. The author would like to thank Sam Braunstein for comments.


\begin{thebibliography}{99}
\bibitem{teleNOTE} In this approximate description, we are not accounting for the fact
that the sender station in teleportation needs to have $n$ input qubits besides the $n$ qubits part of the shared ebits. If we account for this, then we saturate the bound with $n/2$, i.e., we need to consider $2n$ for teleportation in Eq.~\eqref{energy2} [and Eq.~\eqref{eqmain}]. Due to the macroscopic
size of the upper bound, we ignore this extra factor 2 in our derivation.
\bibitem{NOTA_pre} Note that we also recover the previous results for massive qubits.
In fact, assume that each local system, $A$ or $B$, has a
symmetrically distributed mass $M$ with zero charge and angular
momentum. Then, we can use $nk \ln2=S_{A}^{\mathrm{th}}\leq
S_{\mathrm{BH}}$ and Eq.~(\ref{BHeq}) to retrieve the result in
Eq.~(\ref{eqmain}) with $D=2R_{\text{S}}$. Using the mass-energy
equivalence, we then transform Eq.~(\ref{eqmain}) in
Eq.~(\ref{energy2}).
\bibitem {NOTA}Note that, in a typical quantum communication protocol, Alice is
assumed to distribute $N$ identical copies of a TMSV state.
Therefore her total entropy is equal to $NS_{A}^{\mathrm{th}}$ and the
spherical bound imposes a more stringent condition on the average number of
photons per copy
\begin{equation}
\bar{n}\leq e^{\pi N^{-1}R_{p}^{2}-1}. \label{Ncopies}%
\end{equation}
Considering that the asymptotic performance for $N\rightarrow+\infty$ is
assumed in the definition of quantum capacities, we may appreciate how the
bound in Eq.~(\ref{Ncopies}) may affect the current results in quantum
information theory. Because one may argue that the various TMSV states can be
prepared and consumed at different times, we weakly assume the condition of
Eq.~(\ref{limitCV}) in our next derivations.
\bibitem{NOTAcomp} In particular, this can be derived from Sec.~II.B of Ref.~\cite{RicFINITE}, by setting
there $\eta =1$ and then $a=b=\mu$. This choice will provide $c=\sqrt{\mu^2-1/4}$ and $\nu = 2\mu - 2c$
which is the added noise of the additive Gaussian channel that is induced by a unit-gain BK teleportation protocol over
a TMSV state with finite variance $\mu$. In this paper we denote  $\nu$ with the symbol $\xi$.
\bibitem{NOTAcomp2} To do this computation, we first note that the channel $\cal{E}_{\eta}^{\mu}$ acting on mode $B$ is equivalent to a thermal-loss channel with
transmissivity $\eta$ and thermal variance $\omega := \eta \xi/(1-\eta) +1/2$. This can be easily proven from the composition $\cal{E}_{\eta}^{\mu} = \cal{E}_{\eta} \circ \cal{I}^{\mu}$,
where $\cal{E}_{\eta}$ is pure-loss with transmissivity $\eta$ and $\cal{I}^{\mu}$ is additive Gaussian with noise $\xi$ [related to $\mu$ via Eq.~\eqref{added}]. Thus, $\cal{E}_{\eta}^{\mu}$  can be 
dilated into a beam-splitter transformation $S_{BC}^{\eta}$ mixing the input mode $B$ with an environmental mode $C$ in a thermal state $\rho_{C}^{\omega}$ with variance $\omega$. For $\cal{E}_{\eta}$, we can do the same dilation but into a vacuum environment. For any $\rho_{AB}$, we may therefore write
\begin{align}
    &\left\Vert \mathcal{I_{A}}\otimes\mathcal{E}_{\eta}(\rho_{AB}
)-\mathcal{I}_{A}\otimes\mathcal{E}_{\eta}^{\mu}(\rho_{AB})\right\Vert _{1} \nonumber \\
& \le \left\Vert \mathcal{I}_{A}\otimes S_{BC}^{\eta}(\rho_{AB}
\otimes \left\vert 0 \right\rangle_{C} \left\langle 0 \right\vert)-
\mathcal{I}_{A}\otimes S_{BC}^{\eta}(\rho_{AB}
\otimes \rho_{C}^{\omega}) \right\Vert_{1} \nonumber \\
& \le \left\Vert \rho_{AB}
\otimes \left\vert 0 \right\rangle_{C} \left\langle 0 \right\vert - \rho_{AB}
\otimes \rho_{C}^{\omega} \right\Vert _{1} \nonumber \\
& \le \left\Vert  \left\vert 0 \right\rangle_{C} \left\langle 0 \right\vert - \rho_{C}^{\omega} \right\Vert _{1} \nonumber \\
& \overset{(*)}{\le} 2 \sqrt{1- F(\left\vert 0 \right\rangle_{C} \left\langle 0 \right\vert, \rho_{C}^{\omega})^{2}} 
= 2 \sqrt{1- (\omega+1/2)^{-1}} \nonumber \\
& =2\sqrt{\frac{\eta\xi}{\eta\xi+1-\eta}}. \nonumber
\end{align}
Here we exploited several properties of the trace distance, such as its decrease under partial trace and data processing. Then we also used $\left\Vert \rho - \sigma \right\Vert_{1} \le 2 \sqrt{1-F(\rho,\sigma)^{2}}$ in $(*)$ and computed the fidelity. Because the result holds for any state $\rho_{AB}$, it also holds for the supremum in Eq.~\eqref{defDIAM} and, therefore, for the diamond distance.
\bibitem{NOTApeeling} This can be shown by performing the peeling technique described in Eqs.~(103) and~(104) of Ref.~\cite{PLOB}.
\end{thebibliography}
\end{document}